\documentclass[nofootinbib,twocolumn,preprintnumbers]{revtex4-1}
\usepackage{amsmath} 
\usepackage{longtable}
\usepackage{graphicx}
\usepackage{color}
\usepackage{amssymb}
\input{epsf}

\newcommand\f[2]{\frac{#1}{#2}} 
\def\beq{\begin{equation}} 
\def\eeq{\end{equation}} 
\def\to{\rightarrow} 
 
\def\beeq{\begin{eqnarray}}
\def\eeeq{\end{eqnarray}}
\def\sh{\hat{s}}
\def\ca{{\cal A}}

\def\eqn#1{eq.~(\ref{#1})}

\begin{document} 

\begin{titlepage}
\renewcommand{\thefootnote}{\fnsymbol{footnote}}
\hfill SLAC--PUB--16256\quad CALT--TH--2015--018
\quad FR--PHENO--2015--003\\
\vskip0.8cm

\begin{center}

{\Large \bf Interference effects in the $H(\to \gamma\gamma) + 2$ jets channel at the LHC}
\end{center}
\par \vspace{2mm}
\begin{center}
{F. Coradeschi,$^{a,b}$}\footnote{coradeschi@fi.infn.it}\ \ 
{D. de Florian,$^b$}\footnote{deflo@df.uba.ar}\ \ 
{L. J. Dixon,$^{c,d}$}\footnote{lance@slac.stanford.edu}\ \ 
{N. Fidanza,$^b$}\footnote{nfidanza@df.uba.ar}\ \ 

{S. H{\"o}che,$^c$}\footnote{shoeche@slac.stanford.edu}\ \ 
{H. Ita,$^e$}\footnote{harald.ita@physik.uni-freiburg.de}\ \ 
{Y. Li,$^c$}\footnote{yli@slac.stanford.edu}\ \ 
{J. Mazzitelli$^{\,b}$}\footnote{jmazzi@df.uba.ar}

\vspace{5mm}

$^a$ Dipartimento di Fisica e Astronomia, Universit\`a di Firenze e \\ 
INFN, Sezione di Firenze, Via G. Sansone 1, Sesto F.no (FI), Italia

$^b$ Departamento de F\'isica, FCEyN, Universidad de Buenos Aires, \\
(1428) Pabell\'on 1, Ciudad Universitaria, Capital Federal, Argentina

$^c$ SLAC National Accelerator Laboratory, Stanford University, Stanford, CA 94309, USA

$^d$ Walter Burke Institute for Theoretical Physics,
Caltech, Pasadena, CA 91125, USA\\

$^e$ Physikalisches Institut, Albert-Ludwigs-Universit{\"a}t Freiburg, D–79104 Freiburg, Germany
\\

\vspace{5mm}

\end{center}

\begin{center} {\large \bf Abstract} \end{center}
\begin{quote}
We compute the interference between the resonant process $pp\to H(\to\gamma\gamma)+2\text{ jets}$ and the corresponding continuum background at leading order in QCD. For the Higgs signal, we include gluon fusion (GF) and vector boson fusion (VBF) production channels, while for the background we consider all tree-level contributions, including pure EW effects (${\cal O}(\alpha_{QED}^4)$) and QCD contributions (${\cal O}(\alpha_{QED}^2 \alpha_{s}^2)$), plus the loop-induced gluon-initiated process.  After convolution with the experimental mass resolution, the main effect of the interference is to shift the position of the mass peak, as in the inclusive GF case studied previously.  The apparent mass shift is small in magnitude but strongly dependent on the Higgs width, potentially allowing for a measurement of, or bound on, the width itself.  In the $H(\to\gamma\gamma)+2\text{ jets}$ channel, the VBF and GF contributions generate shifts of opposite signs which largely cancel, depending on the sets of cuts used, to as little as 5~MeV (toward a lower Higgs mass).  The small magnitude of the shift makes this channel a good reference mass for measuring the inclusive mass shift of around 60~MeV in the Standard Model.
\end{quote}

\vspace{1cm}

\end{titlepage}

\setcounter{footnote}{1}
\renewcommand{\thefootnote}{\fnsymbol{footnote}}


\section*{Introduction}

In 2012, the ATLAS and CMS collaborations at the Large Hadron Collider (LHC) observed a new particle whose measured properties are, so far, compatible with the Standard Model (SM) Higgs boson~\cite{Aad:2012tfa,Chatrchyan:2012ufa}.  It is now important to study the new particle's properties as accurately as possible, in order to unveil any possible deviations from the predictions of the SM.  

The observed resonance in the diphoton invariant mass at the LHC was one of the main discovery channels, and it provides a very clean signature for probing Higgs properties.  In the inclusive case, i.e.~the resonant process $pp\to H(\to\gamma\gamma) + X$, the main contribution to the signal cross section comes from the Gluon Fusion (GF) mechanism, while the corresponding background is driven by the partonic process $q\bar{q}\to\gamma\gamma$ and its higher order QCD corrections.  The next most prominent production process is Vector Boson Fusion (VBF). It has a smaller rate, but the signal to background ratio in the diphoton channel can be larger after suitable cuts on the two additional jets present in VBF.

The other principal Higgs discovery channel was $pp\to H(\to ZZ^* \to 4~{\rm leptons}) + X$.  The diphoton and $ZZ^*$ mode are also the only decay modes from which a precise measurement of the Higgs boson mass has been obtained.  A recent combined measurement from ATLAS and CMS~\cite{Aad:2015zhl} has determined the masses in the two modes to be,
\begin{eqnarray}
m_H^{\gamma\gamma} &=& 125.07 \pm 0.25\,({\rm stat})\pm 0.14\,({\rm syst})~{\rm GeV}, \label{mHyy}\\
m_H^{ZZ^*} &=& 125.15 \pm 0.37\,({\rm stat})\pm 0.15\,({\rm syst})~{\rm GeV}, \label{mHZZ}
\end{eqnarray}
yielding a mass difference of,
\begin{equation}
m_H^{\gamma\gamma} - m_H^{ZZ^*} = - 80 \pm 490~{\rm MeV}, \label{DeltamHyyZZ}
\end{equation}
neglecting any correlations between the systematic errors.
Because these measurements are currently statistically limited, they should improve significantly in the next run of the LHC.

The effect of the interference between signal and background in the inclusive $\gamma\gamma$ channel has been studied in Refs.~\cite{Dicus:1987fk,Dixon:2003yb,Martin:2012xc,deFlorian:2013psa,Martin:2013ula,Dixon:2013haa}. In particular, as first shown in Ref.~\cite{Martin:2012xc}, the main effect of the interference, after convolution with the broad ($\sim\,1$~GeV) experimental diphoton mass resolution, is to produce a shift in the diphoton mass peak towards lower invariant masses.  At leading order in $\alpha_s$ (LO), this shift is of the order of $100$~MeV~\cite{Martin:2012xc,deFlorian:2013psa,Martin:2013ula}.  Including the dominant next-to-leading order (NLO) contributions, it declines to around 60--70~MeV~\cite{Dixon:2013haa}.

As was pointed out in Ref.~\cite{Dixon:2013haa}, this apparent mass shift could also be used to bound the value of the Higgs width.  In the SM, the width of the Higgs boson is $\Gamma_{\text{SM}} = 4.07$ MeV, far too narrow to observe directly.  In a lineshape model that preserves the signal yields to SM states, the mass shift scales like the square-root of the width, $\delta m_H \propto \sqrt{\Gamma/\Gamma_{\text{SM}}}$.  The mass shift in the $ZZ^*$ channel is negligible compared to that in $\gamma\gamma$~\cite{Kauer:2012hd,Dixon:2013haa}.  Thus, the measurement~(\ref{DeltamHyyZZ}) already implies a bound at $2\sigma$ on the Higgs boson width of order
\begin{equation}
\frac{\Gamma}{\Gamma_{\text{SM}}} < 250,
\label{currentGammabound}
\end{equation}
or around 1~GeV, better than the 2.4~GeV that has been achieved by a direct lineshape measurement~\cite{Khachatryan:2014ira}.

There are other indirect approaches to bounding or measuring the Higgs boson width. For example, it has been proposed to measure the yield of $ZZ$ or $WW$ boson pairs at high invariant mass~\cite{Caola:2013yja,Campbell:2013una,Campbell:2013wga,Campbell:2015vwa}, using the fact that unitarity cancellation involving the SM Higgs for $t\bar{t} \rightarrow ZZ$ or $t\bar{t} \rightarrow WW$ ---which takes place inside the loop for $gg \rightarrow ZZ$ or $WW$, as described in Ref.~\cite{Kauer:2012hd} --- is disrupted in the lineshape model mentioned above. This is a powerful method, already
leading to bounds from CMS~\cite{Khachatryan:2014iha} and ATLAS~\cite{Aad:2015xua} that are of order 
$\Gamma/\Gamma_{\text{SM}} < 4.5$, much smaller than \eqn{currentGammabound}.  However, this method is also
considerably more model-dependent.  As pointed out for example in Refs.~\cite{Englert:2014aca,Englert:2014ffa,Logan:2014ppa}, it can be circumvented by form-factor effects or combined unitarity cancellations from other, yet unobserved Higgs bosons.  It also appears difficult at present to push this method all the way down to $\Gamma/\Gamma_{\text{SM}} \approx 1$, due to theoretical uncertainties on the $ZZ$ and $WW$ continuum backgrounds.

In contrast, the mass-shift method operates very close to the Higgs resonance, produces a distinctive signature, and is not affected by other physics that might take place at higher energy scales.  On the other hand, it will be challenging to reduce the uncertainty in~\eqn{DeltamHyyZZ} by another order of magnitude, in order to probe widths of order the Standard Model prediction.  While the uncertainty currently is dominated by statistics, at some point systematic uncertainties will become important.  In particular, the systematic uncertainty on $m_H^{\gamma\gamma}$ is determined largely by the photon energy calibration, whereas the uncertainty for $m_H^{ZZ^*}$ is a combination of electron and muon calibrations.  The momenta of muons are determined from tracking, while photon and electron energies are derived from the electromagnetic calorimeters.  Nevertheless, electron and photon response is not identical, and the difference plays a large role in using the $Z$ mass in $Z\to e^+e^-$ to calibrate $m_H^{\gamma\gamma}$~\cite{Aad:2014aba,Khachatryan:2014ira}.  Finding another reference mass besides $m_H^{ZZ^*}$, with photons in the final state, might lead to reduced systematic uncertainties.  In Ref.~\cite{Dixon:2013haa} it was proposed to use a subsample of the inclusive GF $\gamma\gamma$ sample with nonzero Higgs transverse momentum $p_{T,H}$, taking advantange of a strong dependence of the mass shift on $p_{T,H}$~\cite{Martin:2013ula}.  However, this dependence is also difficult to predict very precisely theoretically.

In this paper we propose using another $\gamma\gamma$ sample, in which the two photons are produced in association with two jets.  Although this process is relatively rare, so is the background, making it possible to obtain reasonable statistical uncertainties on the position of the mass peak in this channel, despite the lower number of events.  The production of a Higgs in association with two jets is characteristic of the Vector Boson Fusion (VBF) production mechanism.  While, in general terms, VBF is subdominant with respect to GF, it has a very different kinematical signature and can be selected through an appropriate choice of the experimental cuts.  From a theoretical point of view, the VBF production mechanism has the additional advantage that perturbative corrections are much smaller than for GF (see e.g.~Ref.~\cite{Bolzoni:2010xr}).  In the following, we will study the effect of the signal-background interference for both the GF and VBF production mechanisms.\footnote{Recently, a similar study has been performed for VBF Higgs production in $e^+e^-$ annihilation~\cite{Liebler:2015eja}, where there is no competition from GF, and a shift of order $100$~MeV was found.}  By adjusting the cuts on the associated jets, we can use a cancellation between GF and VBF mass shifts to minimize the mass shift in this sample, making it an excellent reference mass for studies at high LHC luminosity, instead of or in addition to the $ZZ^*$ reference mass in \eqn{DeltamHyyZZ}.


\section*{Overview of the calculation}

The interference of the resonant production process $i_1i_2 \to H ( \to \gamma \gamma ) + 2 j$ with the continuum background $i_1i_2 \to \gamma\gamma + 2j $ can be expressed at the level of the partonic cross section as:
\begin{eqnarray}
\delta\hat{\sigma}_{i_1i_2\to H(\to\gamma\gamma)+2j} 
\!&=&\!\!\! 
 \nonumber\\
&& \hskip-2cm
-2 (\sh-m_H^2)\,\, {\text{Re} \left( \ca_{i_1i_2\to H+2j} \ca_{H\to\gamma\gamma} 
                          \ca_{\rm cont}^* \right) 
        \over (\sh - m_H^2)^2 + m_H^2 \Gamma^2 }
\nonumber\\
&& \hskip-2cm
-2 m_H \Gamma { \text{Im} \left( \ca_{i_1i_2\to H+2j} \ca_{H\to\gamma\gamma} 
                          \ca_{\rm cont}^* \right)
        \over (\sh - m_H^2)^2 + m_H^2 \Gamma^2 } \,,
\label{intpartonic}
\end{eqnarray}
where $\sh = s_{i_1i_2}$ is the square of the center-of-mass energy for the two incoming partons $i_1$ and $i_2$, ``$2j$'' stands for the two outgoing partons, and $m_H$ and $\Gamma$ are the Higgs mass and decay width, respectively\footnote{The details of the implementation of the lineshape \cite{Goria:2011wa} have a very small effect on the light, narrow Higgs discussed in this note, so we can rely on a naive Breit-Wigner prescription.}.

In practice, the Higgs resonance is very narrow, much narrower than the experimental resolution. As pointed out in Refs.~\cite{Dicus:1987fk,Dixon:2003yb}, this implies that the first term in \eqn{intpartonic}, arising from the real part of the Breit-Wigner --- which is odd in $\sh$ around $m_H$ --- is strongly suppressed by the integral across the resonance. (This is true provided that all $\sh$-dependent functions vary slowly across the resonance, which is the case.)  However, as shown by Martin~\cite{Martin:2012xc}, the experimental smearing does leave behind a quantifiable effect on the position of the diphoton invariant mass peak, \emph{shifting} it to lower masses by ${\cal O}(100\,\text{ MeV})$ at LO.
The reason for this shift is the antisymmetric nature of the interference: the differential cross section gets slightly enhanced (or suppressed, depending on the sign of the interference term) below, and suppressed (enhanced) above $m_H^2$, causing the peak of the smeared invariant-mass distribution to move lower (higher).  The effect is larger than one might naively expect because the real part of the Breit-Wigner has a large tail, proportional to $1/\hat{s}$, which gets enhanced by the experimental smearing, in a manner which is roughly proportional to the experimental resolution. In the end, the observed shift arises from an interplay between a theoretical and an experimental effect.

In contrast, the imaginary part of the Breit-Wigner (the second term in \eqn{intpartonic}) has the same dependence on $\sh$ as the signal contribution and can, in principle, give a significant contribution to the cross section. However, this term turns out to be quantitatively suppressed because it requires a relative phase between signal and background amplitudes~\cite{Dixon:2003yb}.  We do not consider it in the following, focusing instead on the mass shift.

\begin{figure}
\begin{center}
\includegraphics[width=0.35\textwidth]{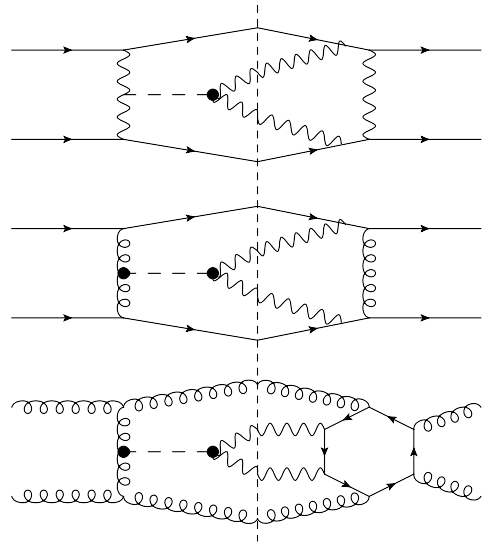}
\caption{Examples of contributing Feynman diagrams. The vertical dotted line separates the Higgs signal (left) from continuum background (right). The top diagram shows the VBF signal and EW background contributions; the middle, the GF signal with tree level QCD mediated background; the bottom, the gluon-initiated signal, with the corresponding loop-induced LO background.}
\label{fig:diagrams}
\end{center}
\end{figure}

In the present article we consider both GF and VBF Higgs production mechanisms for the resonant process $pp\to H(\to\gamma\gamma)+2\text{ jets}$ at LO. For the background, we include at tree level the pure electroweak (EW) effects (${\cal O}(\alpha_{QED}^4)$) and the contributions from QCD (${\cal O}(\alpha_{QED}^2 \alpha_{s}^2)$). In addition, the loop-induced gluon-initiated process $g g \to\gamma\gamma + g g$ was computed. Although of higher order, it is enhanced by the large gluon luminosity at the LHC, and so it is worth checking how large an effect it produces. Examples of the relevant Feynman diagrams can be seen in figure~\ref{fig:diagrams}.

It is possible for the GF and VBF contributions to interfere with each other, in both signal and background.
For the Higgs signal this effect has been studied and found to be very small~\cite{Andersen:2006ag,Andersen:2007mp,Ciccolini:2007ec,Bredenstein:2008tm}.  At tree level, the absence of such VBF/GF interference is due to the different color quantum numbers exchanged in the $t$-channel, singlet versus octet, as can be seen in figure~\ref{fig:diagrams}.  (In the case of $ZZ$ fusion, if the two quarks are identical, interference is allowed but it is highly suppressed kinematically~\cite{Andersen:2006ag}.)  At one loop, the color restriction is relaxed but the interference is still very small~\cite{Andersen:2007mp,Ciccolini:2007ec,Bredenstein:2008tm}.  Similar considerations apply to VBF/GF interference in the continuum background, and to the signal-background interference.  Although we include all terms in the amplitude sums in \eqn{intpartonic}, in practice it is quite accurate to speak of the VBF and GF contributions separately, and we will exhibit results for these individual contributions below.

All of our results were obtained by two independent calculations. In one approach, the relevant Feynman diagrams were obtained with the help of the Mathematica package {\sc FeynArts}~\cite{Hahn:2000kx}. Then, the corresponding analytical expressions were found using a customized version of the package {\sc FormCalc}~\cite{Hahn:1998yk}.  Finally, a dedicated Fortran code was used to assemble the contributions from the various channels, convolute them with the parton density functions (PDFs), and integrate them numerically over the final-state phase-space.

In the other approach, the {\sc SHERPA} event generator~\cite{Gleisberg:2008ta,Hoche:2014kca} with its internal matrix element generator {\sc COMIX}~\cite{Gleisberg:2008fv} was used to compute all tree-level amplitudes, which were cross-checked with {\sc MADGRAPH5}~\cite{Alwall:2011uj}.  For the partonic channel of $gg\to H \to gg \gamma\gamma$, the background continuum process starts at one loop; its matrix element was provided by the {\sc BlackHat} library~\citep{Berger:2008sj,Berger:2009ep,Berger:2010vm,Berger:2010zx}.

Both approaches are in perfect agreement. Results for the various separate signal and background contributions were also cross-checked against the {\sc MCFM} code~\cite{MCFM,Berger:2004pca}.

We used the MSTW2008 LO PDF set~\cite{Martin:2009iq}, along with its corresponding value of $\alpha_{s}(M_Z)$, with the factorization and renormalization scales set equal to the Higgs mass ($\mu_F=\mu_R= m_H$), and 5 massless quark flavours.  For the GF production mechanism we considered the $m_t\to\infty$ limit, using the effective Lagrangian approach for the $ggH$ coupling, which provides a good approximation even for large di-jet invariant masses, as long as the jet transverse momenta are small compared with $m_t$~\cite{DelDuca:2003ba}.
For the decay into two photons, both top quark and $W$ boson loop contributions are sizeable; the former was once again treated in the large $m_t$ limit, while for the latter we used $m_W=80.385\text{ GeV}$. 
For the Higgs boson mass and width we used $m_H=125\text{ GeV}$ and $\Gamma=4.07\text{ MeV}$, and we set $\alpha_{QED}=1/137$ and the Higgs vacuum expectation value to $v=246$~GeV. We performed all the calculations for a collider energy of 14~TeV.


\section*{Mass shift and asymmetry}

All plots presented in this section correspond to the following set of selection cuts: an asymmetric cut on the transverse momenta of the photons, $p_{T, \gamma}^{\text{hard\,(soft)}} > 40~(30)$~GeV; an asymmetric cut on the transverse momenta of the jets, $p_{T, j}^{\text{hard\,(soft)}} > 40~(25)$~GeV; a symmetric constraint on the photon and jet pseudorapidities, $|\eta_\gamma | < 2.5$ and $|\eta_j | < 4.5$; and standard isolation cuts for the photons and jets, requesting $R_{\gamma\gamma}, R_{\gamma j}, R_{jj} > 0.4$, where $R=\sqrt{\Delta \phi^2+\Delta \eta^2}$.
We also impose a cut in the invariant mass of the dijet system of  $M_{jj}>400\text{ GeV}$.  We consider a range of possible cuts on the difference in pseudorapidities between the jets ($\Delta\eta_{jj}$). The cut $|\Delta\eta_{jj}| > |\Delta\eta_{jj}|_{\text{min}}$, in particular, enhances the contribution of the VBF production channel, which characteristically yields two very forward jets.  Finally, we request a minimum transverse momentum for the Higgs/diphoton pair,
$p_{T,H} \equiv | \vec{p}_{T,\gamma}^{\text{\,hard}} + \vec{p}_{T,\gamma}^{\text{\,soft}} |$.
We impose $p_{T,H} > p_{T,H}^{\text{min}}$, with $p_{T,H}^{\text{min}}$ varied in the range from $0$ to $160$ GeV.

In order to simulate the detector resolution, we convolute the cross section with a Gaussian function in the diphoton invariant mass, following the procedure of Ref.~\cite{Martin:2012xc}. We used several test mass resolution widths $\sigma_{\text{MR}}$ of the order of 1 GeV. The precise value of the apparent mass shift $\delta m_H$ is roughly proportional to both the width of the Gaussian, and the absolute magnitude and sign of the interference. For the precise definition of the procedure used to obtain the numerical value of the mass shift, we refer to Ref. \cite{Dixon:2013haa}. In the following results we use a mass resolution of $\sigma_{\text{MR}} = 1.7$ GeV as the benchmark value.

\begin{figure}
\begin{center}
\hskip-0.4cm \includegraphics[width=0.48\textwidth]{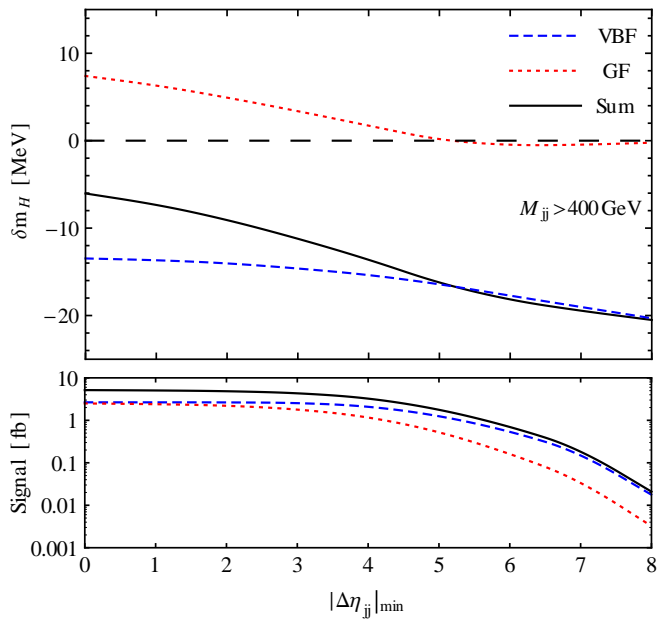}
\caption{Top: Plot of mass shift $\delta m_H$ for different values of $|\Delta\eta_{jj}|_{\text{min}}$. The dashed blue line represents the contribution from the VBF mechanism alone, the dotted red line shows GF only, and the solid black line displays the total shift of the Higgs invariant mass peak. Bottom: Total integrated signal cross section, also separated into VBF and GF contributions for the same cuts. No cut on $p_{T,H}^{\text{min}}$ was applied, and an additional cut was set of $M_{jj}>400\text{ GeV}$.}
\label{fig:eta}
\end{center}
\end{figure}

In figure \ref{fig:eta} we show the values of the apparent mass shift $\delta m_H$ obtained for different cuts on $|\Delta\eta_{jj}|$. We present the contributions from VBF and GF separately, as well as the total shift. At the bottom of the plot we show the total integrated signal, also separated into VBF and GF contributions for the same cuts. For this plot no cut in $p_{T,H}$ was applied, and we considered only events with $M_{jj}>400\text{ GeV}$. When no cut in $|\Delta\eta_{jj}|$ is applied, the shift in the Higgs invariant mass peak position produced by these two main production mechanisms is of the same magnitude, but of opposite sign; hence we observe a partial cancellation between them, with a net shift of around $-6\text{ MeV}$. As the value of $|\Delta\eta_{jj}|_{\text{min}}$ is increased, VBF becomes the dominant contribution, and GF turns negligible, leading to a shift of around $20 \text{ MeV}$ toward lower masses.

\begin{figure}
\begin{center}
\hskip-0.4cm \includegraphics[width=0.48\textwidth]{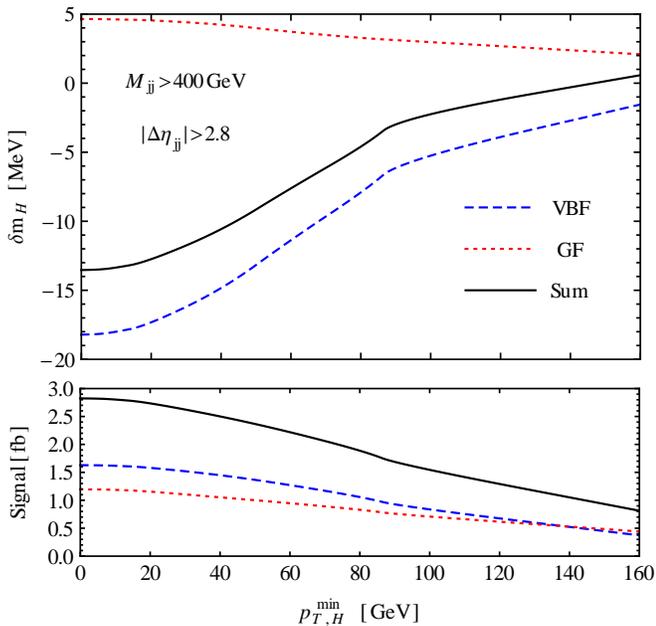}
\caption{Top: Plot of mass shift $\delta m_H$ for different values of $p_{T,H}^{\text{min}}$ for VBF, GF and total contributions.
The curves are labeled as in figure~\ref{fig:eta}.  Bottom:  Total integrated signal, also separated into VBF and GF contributions for the same cuts.  The following additional cuts were applied: $M_{jj}>400\text{ GeV}$ and $|\Delta\eta_{jj}|>2.8$.}
\label{fig:ptH}
\end{center}
\end{figure}

Next we study the dependence of the mass shift on $p_{T,H}^{\text{min}}$. In figure \ref{fig:ptH} we present the mass shift and the signal cross section for a range of $p_{T,H}^{\text{min}}$ between $0 \text{ GeV}$ and $160 \text{ GeV}$. The curves are labeled in the same way as in figure~\ref{fig:eta}.  Once again, both production mechanisms contribute to the shift in invariant mass with opposite signs. For this plot, we applied the additional cuts of $M_{jj}>400\text{ GeV}$ and $|\Delta\eta_{jj}|>2.8$, enhancing in this way the VBF contributions.  However, at higher $p_{T,H}^{\text{min}}$, GF becomes as important as VBF.

\begin{figure}
\begin{center}
\hskip-0.4cm \includegraphics[width=0.48\textwidth]{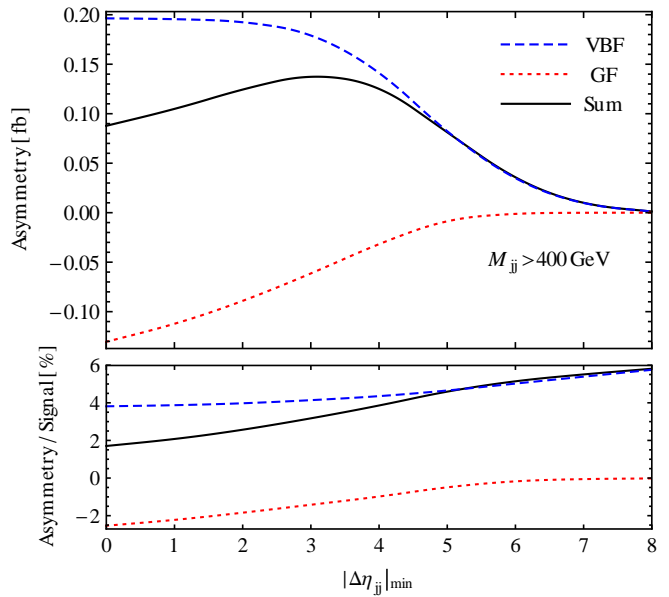}
\caption{Plot of asymmetry ${\cal A}$ for different values of $|\Delta\eta_{jj}|_{\text{min}}$, for VBF, GF and total contributions.  The curves are labeled as in figure~\ref{fig:eta}.  Bottom:  Plot of the ratio between the asymmetry and the total integrated signal, also separated into VBF and GF contributions for the same cuts. No cut in $p_{T,H}^{\text{min}}$ was applied, and an additional cut was set of $M_{jj}>400\text{ GeV}$.}
\label{fig:asym_eta}
\end{center}
\end{figure}

\begin{figure}
\begin{center}
\hskip-0.4cm \includegraphics[width=0.48\textwidth]{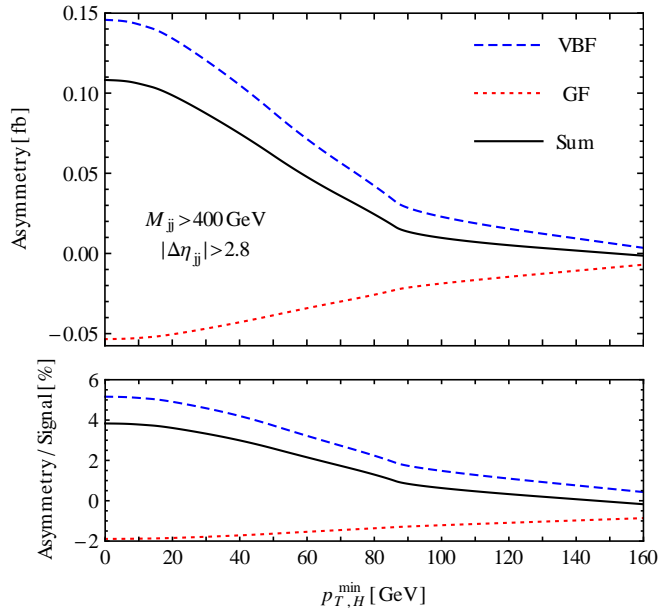}
\caption{Plot of asymmetry ${\cal A}$ for different values of $p_{T,H}^{\text{min}}$, for VBF, GF and total contributions.  The curves are labeled as in figure~\ref{fig:eta}.  Bottom:  Plot of the ratio between the asymmetry and the total integrated signal, also separated into VBF and GF contributions for the same cuts.  Additional cuts were applied of $M_{jj}>400\text{ GeV}$ and $|\Delta\eta_{jj}|>2.8$.}
\label{fig:asym_ptH}
\end{center}
\end{figure}

In order to portray the effect of the interference terms, we study the asymmetry ${\cal A}$, defined as the difference between the interference contribution to the hadronic cross section ($\delta\sigma$) for diphoton masses smaller than the Higgs mass, minus that for larger diphoton masses. More specifically, we define:
\beq
{\cal A} = \int_{115 \text{ GeV}}^{125 \text{ GeV}}
\f{d(\delta\sigma)}{dM_{\gamma\gamma}}dM_{\gamma\gamma}
-
\int_{125 \text{ GeV}}^{135 \text{ GeV}}
\f{d(\delta\sigma)}{dM_{\gamma\gamma}}dM_{\gamma\gamma}\,.
\eeq
This quantity serves as a theoretical proxy for the mass shift; no Gaussian smearing has been applied. We plot the asymmetry for different values of $|\Delta\eta_{jj}|_{\text{min}}$ in figure \ref{fig:asym_eta} and for different values of $p_{T,H}^{\text{min}}$ in figure \ref{fig:asym_ptH}.  Once again we see how the contributions from VBF and GF are of opposite signs, and therefore their effect cancels out partially. At the bottom of both figures we show the ratio between the asymmetry and the total integrated signal.  (Note that the denominator is the Higgs signal integrated across the resonance, not the continuum background in the given broad range of $M_{\gamma\gamma}$.)

\begin{figure}
\begin{center}
\hskip-0.4cm \includegraphics[width=0.5\textwidth]{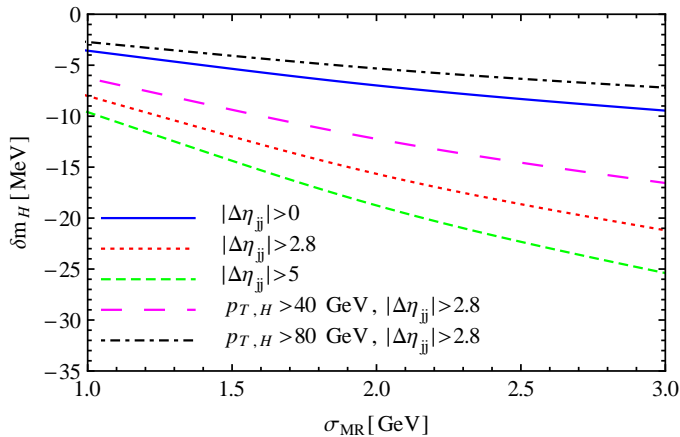}
\caption{Plot of mass shift for different values of mass resolution $\sigma_{\text{MR}}$. All the results have a cut of $M_{jj}>400\text{ GeV}$. The solid blue line shows the results with no additional cuts; the dotted red line for a cut of $|\Delta \eta_{jj}|>2.8$; the dashed green line for $|\Delta \eta_{jj}|>5$; the long dashed magenta line for $p_{T,H}^{\text{min}}>40$ GeV and $|\Delta \eta_{jj}|>2.8$; and the dot-dashed black line for $p_{T,H}^{\text{min}}>80$ GeV and $|\Delta \eta_{jj}|>2.8$.}
\label{fig:resolution}
\end{center}
\end{figure}

We also studied the dependence of the mass shift on the width of the Gaussian used to simulate the experimental mass resolution of the detector $\sigma_{\text{MR}}$. Figure~\ref{fig:resolution} shows that $\delta m_H$ increases with $\sigma_{\text{MR}}$ in a roughly linear way, for five different choices of cuts.

\section*{Bounding the Higgs Width}

As we have seen in the previous section, the shift in the Higgs invariant mass peak in $pp\to H(\to\gamma\gamma)+2\text{ jets}\,+\,X$ is considerably smaller than in the inclusive channel $pp\to H(\to\gamma\gamma)\,+\,X$.  For appropriate cuts it can be almost zero.  This makes it useful as a reference mass for experimental measurement of the mass difference,
\begin{equation}
\Delta m_H^{\gamma\gamma} \equiv \delta m_H^{\gamma\gamma,\,\text{incl}} - \delta m_H^{\gamma\gamma,\,\text{VBF}} \,,
\label{BigDeltaDef}
\end{equation}
where $\delta m_H^{\gamma\gamma,\,\text{incl}}$ is the mass shift in the inclusive channel, as computed at NLO in Ref.~\cite{Dixon:2013haa}, and $\delta m_H^{\gamma\gamma,\,\text{VBF}}$ is the quantity computed in this paper.  In computing $\delta m_H^{\gamma\gamma,\,\text{VBF}}$ for use in \eqn{BigDeltaDef} we impose the basic photon and jet $p_T$ and $\eta$ cuts, and $M_{jj} > 400$~GeV, but no additional cuts on $p_{T,H}$ or $\Delta\eta_{jj}$.  This choice of cuts results in a small reference mass shift and a relatively large rate with which to measure it.

All the calculations we have presented so far were carried out by setting the Higgs width $\Gamma$ to the one predicted by the SM: $\Gamma_{\text{SM}} = 4.07$ MeV.  In this section we use the lineshape model of Ref.~\cite{Dixon:2013haa} to compute the mass shift for a variable width $\Gamma$, in a way that is relatively independent of the new physics that increases $\Gamma$ from the SM value.  To be consistent with the Higgs signal strength measurements already made by the LHC, if the value of the Higgs width is varied its couplings must also be modified, in order to prevent the total cross section from suffering large variations.  We assume a model in which the couplings of the Higgs boson to the top quark and the massive weak bosons deviate from the SM predictions by real factors $c_t$ and $c_V$, respectively.  This generates a variation in the effective coupling of the Higgs to gluons and photons by real factors $c_g$ and $c_\gamma$.  We adjust $\Gamma$ to maintain the Higgs signal strength near the SM value.  For example, for the $\gamma\gamma$ channel we have, integrating over the resonance in the narrow-width approximation~\cite{Dixon:2013haa},
\begin{equation}
\frac{c_{g\gamma}^2 S}{m_H\Gamma}+c_{g\gamma} I 
= \left( \frac{S}{m_H\Gamma_{\text{SM}}} + I \right)
 \mu_{\text{GF}} \,,
\label{constyieldGF}
\end{equation}
where $c_{g\gamma} \equiv c_{g}c_\gamma$, $S$ is the SM Higgs signal cross section, and $\mu_{\text{GF}}$
denotes the ratio of the experimental signal strength in $gg\to H\to \gamma\gamma$ to the SM prediction
($\sigma/\sigma^{\rm SM}$).  The interference term $I$ is negligible;
the fractional destructive interference in the SM is 
$m_H\Gamma_{\text{SM}} \, I/S \approx -1.6\%$~\cite{Dixon:2003yb}.

An analogous equation holds for the VBF production of Higgs bosons decaying to $\gamma\gamma$,
\begin{equation}
\frac{c_{V\gamma}^2 S}{m_H\Gamma}
= \frac{S}{m_H\Gamma_{\text{SM}}} \mu_{\text{VBF}} \,,
\label{constyieldVBF}
\end{equation}
where $c_{V\gamma} \equiv c_{V}c_\gamma$, and $\mu_{\text{VBF}}$ denotes the ratio of the experimental signal strength to SM prediction in VBF production with decay to $\gamma\gamma$.  We have dropped the corresponding interference term $I$ because it is even smaller in this case.  Neglecting $I$ also in \eqn{constyieldGF}, we see that
$c_{g\gamma}^2/\Gamma = \mu_{\text{GF}}/\Gamma_{\text{SM}}$ and $c_{V\gamma}^2/\Gamma = \mu_{\text{VBF}}/\Gamma_{\text{SM}}$,
whose solution is
\begin{equation}
c_{g\gamma} = \sqrt{\frac{\mu_{\text{GF}}\Gamma}{\Gamma_{\text{SM}}}} \,, \qquad
c_{V\gamma}=\sqrt{\frac{\mu_{\text{VBF}}\Gamma}{\Gamma_{\text{SM}}}} \,.
\label{csol}
\end{equation}
The current experimental values for the signal strengths from ATLAS are $\mu_{\text{VBF}}=0.8 \pm 0.7$ and $\mu_{\text{GF}}=1.32 \pm 0.38$ \cite{Aad:2014eha}, and from CMS $\hat\mu_{\text{VBF}}=1.58 \pm 0.7$ and $\hat\mu_{\text{GF}}=1.12 \pm 0.37$ \cite{Khachatryan:2014ira}, which are compatible with the SM predictions.\footnote{These are not precisely the same quantities as we have defined, since they include information from the $ZZ^*$ final state as well as $\gamma\gamma$.}
We used for our analysis the values $\mu_{\text{GF}}=\mu_{\text{VBF}}=1$.

\begin{figure}
\begin{center}
\hskip-0.4cm \includegraphics[width=0.5\textwidth]{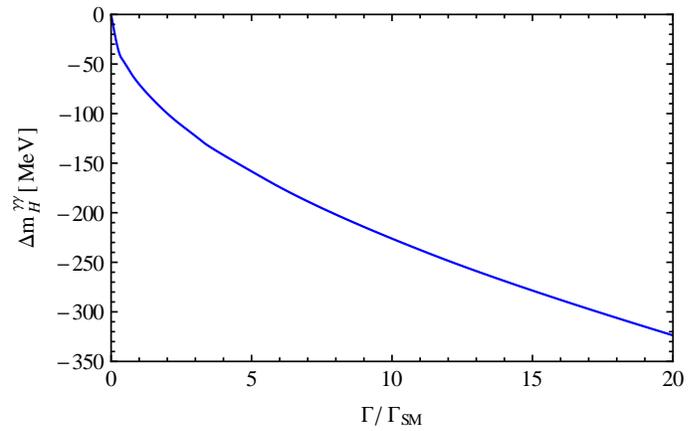}
\caption{Plot of measurable mass shift $\Delta m_H^{\gamma\gamma}$ defined in \eqn{BigDeltaDef}, as a function of
$\Gamma / \Gamma_{\text{SM}}$.}
\label{fig:widths}
\end{center}
\end{figure} 

In figure \ref{fig:widths} we show how the observable $\Delta m_H^{\gamma\gamma}$ depends on the value of the Higgs width.  The dependence is proportional to $\sqrt{\Gamma/\Gamma_{\text{SM}}}$ to a very good accuracy, as dictated by \eqn{csol} and the linearity of the produced shift in $c_{g\gamma}$ or $c_{V\gamma}$ (in the range shown).  It is dominated by the mass shift for the inclusive sample~\cite{Dixon:2013haa}. 


\section*{Conclusions}

In this paper, we studied the interference between the resonant process $pp\to H(\to \gamma\gamma)+\text{2 jets}$ and the corresponding continuum background at the LHC, which generates a shift in the position of the peak in the diphoton invariant-mass spectrum.  A similar shift also occurs for the $\gamma\gamma$ final state in inclusive Higgs production.  The shift is strongly dependent on the Higgs width, in models where the Higgs experimental yields are held constant.
This feature might allow LHC experiments to measure or bound the width.

Both gluon fusion and vector boson fusion production mechanisms contribute to $pp\to H+\text{2 jets}$.  The two mechanisms generate shifts of opposite signs, leading to a large cancellation of this effect.  Depending on the sets of cuts that are used, the net shift can be as small as $5 \text{ MeV}$ toward lower masses.  The small magnitude of this shift makes this channel a good reference mass for measuring the larger mass shift of the inclusive case.
The expected measurable mass shift, defined as the mass difference between the peaks of the $\gamma\gamma+\text{2 jets}+X$ sample and of the inclusive diphoton sample, has been obtained as a function of the Higgs width (figure \ref{fig:widths}).  Our main theoretical assumption was that the couplings of the Higgs rescale by real factors.  We also assumed the same rescaling for the Higgs coupling to gluons as for its coupling to vector boson pairs; this assumption could easily be relaxed, to the degree allowed by current measurements of the relative yields in different channels.

We look forward to more theoretical studies along these lines, as well as future experimental efforts at the LHC to measure or bound this shift, and hence the width of the Higgs boson. 


\section*{Acknowledgments}

This work was supported in part by UBACYT, CONICET, ANPCyT, the Research Executive Agency (REA) of the European Union under the Grant Agreement number PITN-GA-2010-264564 (LHCPhenoNet), the US Department of Energy under contract DE--AC02--76SF00515, the Walter Burke Institute at Caltech, and the Gordon and Betty Moore Foundation through Grant No.~776 to the Caltech Moore Center for Theoretical Cosmology and Physics. LD and DdF thank the Galileo Galilei Institute for hospitality during a portion of this work.


\bibliography{bibliography}

\end{document}